**Short Paper***

# Web-Based File Clustering and Indexing for Mindoro State University

Christine A. Luzon
College of Computer Studies, Mindoro State University, Philippines
christineluzon2020@gmail.com
(corresponding author)

Luisito L. Lacatan
College of Engineering, Laguna University, Philippines
louie1428@gmail.com

Harold Y. Bangalisan
College of Computer Studies, Mindoro State University, Philippines

Jayvee M. Osapdin
College of Computer Studies, Mindoro State University, Philippines

## ABSTRACT

*Purpose* – The Web-Based File Clustering and Indexing for Mindoro State University aim to organize data circulated over the Web into groups/collections to facilitate data availability and access, and at the same time meet user preferences. The main benefits include:

increasing Web information accessibility, understanding users' navigation behavior, improving information retrieval and content delivery on the Web. Web-based file clustering could help in reaching the required documents that the user is searching for.

*Method* – In this paper a novel approach has been introduced for search results clustering that is based on the semantics of the retrieved documents rather than the syntax of the terms in those documents. Data clustering was used to improve the information retrieval from the collection of documents. Data were processed and analyzed using SPSS (version 18) where the instrument was evaluated to test the reliability and validity of the measures used. Evaluation was based on a Likert scale of Excellent, Good, Fair, and Poor as described for the selected quality characteristics.

*Results* – A total of 200 questionnaires were distributed with a return rate of 100%. The questionnaire was tested 0.735 using Cronbach's Alpha Coefficient and considered a reliable instrument. Four quality characteristics were evaluated in this study; Usability, Performance Efficiency, Reliability, and Functionality Suitability.

*Conclusion* - The Web-based file clustering could help in reaching the required documents that the user is searching for. The need for an information retrieval mechanism can only be supported if the document collection is organized into a meaningful structure, which allows part or all the document collection to be browsed at each stage of a search.

*Recommendations* – It is recommended that upon uploading of file it will show the use of the file and where it is originated (department). It is also recommended to create an index to cluster not only the file type but also the content and use of a file. Explore the clustering to a wider scope.

*Practical Implications* – Document clustering provides a structure for organizing large bodies of text for efficient browsing and searching and helps a lot for the Mindoro State University for records/ document processing. Indexing is the best tool to maintain uniqueness of records in a database. Whenever new files or records are created, it can be easily added to the index. This makes it easy to keep documents up-to-date at all times. Grouping documents into two or more categories improves search time and makes life easier for everyone.

## INTRODUCTION

The internet has become the largest data repository facing the problem of information overload. This information explosion has led to a growing challenge for information retrieval systems to efficiently and effectively manage and retrieve the information for an average user. The purpose of file clustering is to store documents electronically and assist the user in effectively navigating, tracing, and organizing the available web documents. The system accepts a query from the user and responds with a set of documents. The system returns both relevant and non-relevant material and a document organization approach is applied to assist the user in finding the relevant information in the retrieved set.

Text document clustering groups similar documents that form a coherent cluster, while documents that are different have separated apart into different clusters. However, the definition of a pair of documents being similar or different is not always clear and normally varies with the actual problem setting. For example, when clustering research papers, two documents are regarded as similar if they share similar thematic topics. When clustering is employed on websites, we are usually more interested in clustering the component pages according to the type of information that is presented on the page. For instance, when dealing with universities' websites, we may want to separate professors' home pages from students' home pages, and pages for courses from pages for research projects. This kind of clustering can benefit further analysis and utilization of the dataset such as information retrieval and information extraction, by grouping similar types of information sources together.

The user will start at the top of the list and follow it down examining the documents one at a time. These approaches are normally based on visualization and presentation of some relationships among the documents and the user's query. Clustering is the method of grouping a group of physical abstract objects into categories of comparable objects.

A cluster could be an assortment of knowledge objects that are kind of like alternatives inside the same cluster and are dissimilar to the objects in other clusters. To enhance the classification task clustering is used as a method to extract information from the unlabeled data. The unlabeled data cluster is mainly used to create a training set. Technology has been improved a lot on World Wide Web. The increasing size and dynamic content of the World Wide Web have created a need for the automated organization of web pages. Document clusters can provide a structure for organizing large bodies of text for efficient browsing and searching.

Generally, the study aims to develop a system of Web-based File clustering and indexing for Mindoro State University. Specifically, the study aimed to store documents electronically and assist the user to effectively navigate, trace and organize the available web documents. The second is to organize data circulated over the Web into groups or collections to facilitate data availability and access and at the same time meet user preferences. Lastly, to provide a structure for organizing large bodies of text for efficient browsing and searching.

## LITERATURE REVIEW

Wishart and Leach apply three different methods of cluster analysis to a data-set consisting of percentage occurrences of 5-syllable sequences throughout 33 passages of Platonic text, in an attempt to analyze Platonic prose rhythm; and Parks applies a hierarchical clustering technique to Purdy's data on the constituent particle composition of recent Bahamian bottom sediment samples. However, sufficient details of the analysis are not always given and motivation for the use of the techniques in the first instance is often omitted.

Premalatha and Natarajan (2009) presented a procreant PSO algorithm for document clustering. This algorithm is a hybrid of Particle Swarm Optimization and Genetic Algorithm, a population-based heuristic search technique, which can be used to solve combinatorial optimization problems, modeled on the concepts of cultural and social rules derived from the analysis of the swarm intelligence (PSO) and also based on crossover and evolution (GA). In standard PSO the non-oscillatory route can quickly cause a particle to stagnate and also, it may prematurely converge on suboptimal solutions that are not even guaranteed to a locally optimal solution. The proposed modification strategy for the PSO algorithm and applied to the document corpus.

## METHODOLOGY

The programming language that was utilized to build up the whole application in Python. The Front end was created utilizing Django, NLTK as one of the libraries for processing of words, and for the backend is Python (Figure 1).

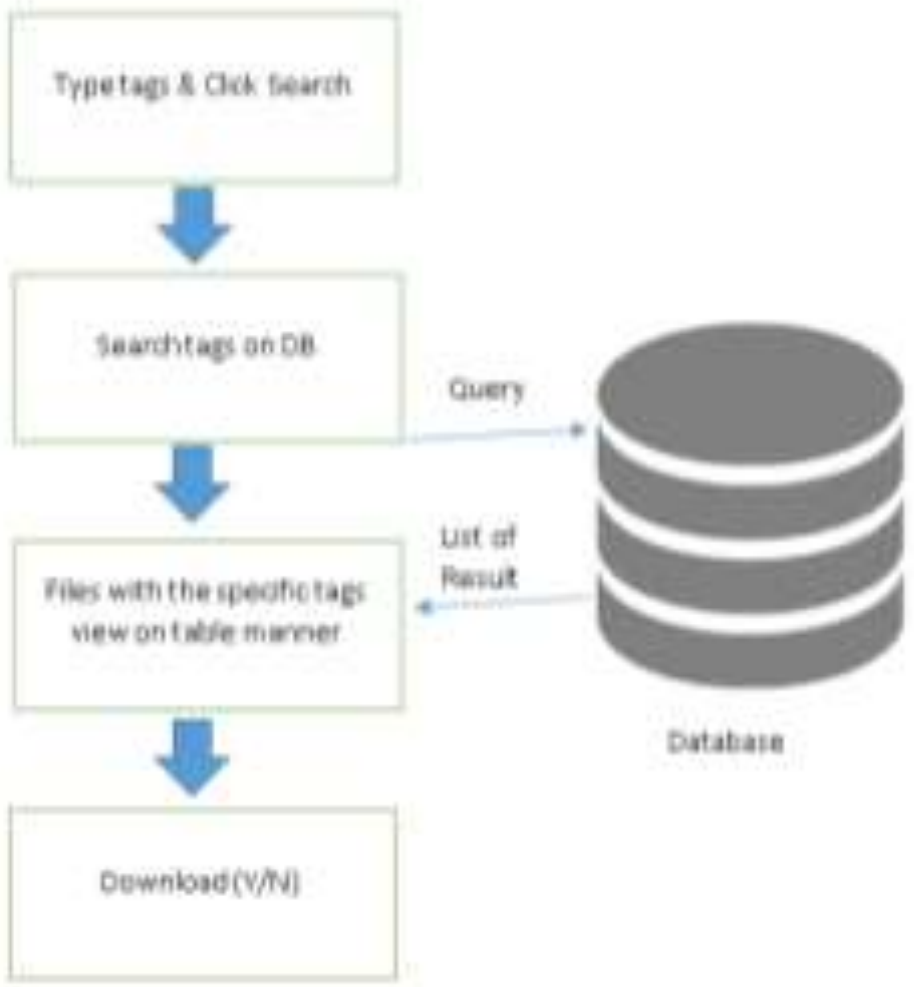

*Figure 1.* Clustering Architecture

Initially, the user will select and upload a file from the collection of documents. The system determines and analyzes the uploaded file (file format). The documents clustering algorithms attempt to group the documents using similarity measures. The stop words are used to eliminate the unwanted words such as before, is, a, an, the, become, then, they, there, that, them, etc. The database contains clustering, file content manipulation, file content tagging, and Optical character recognition.

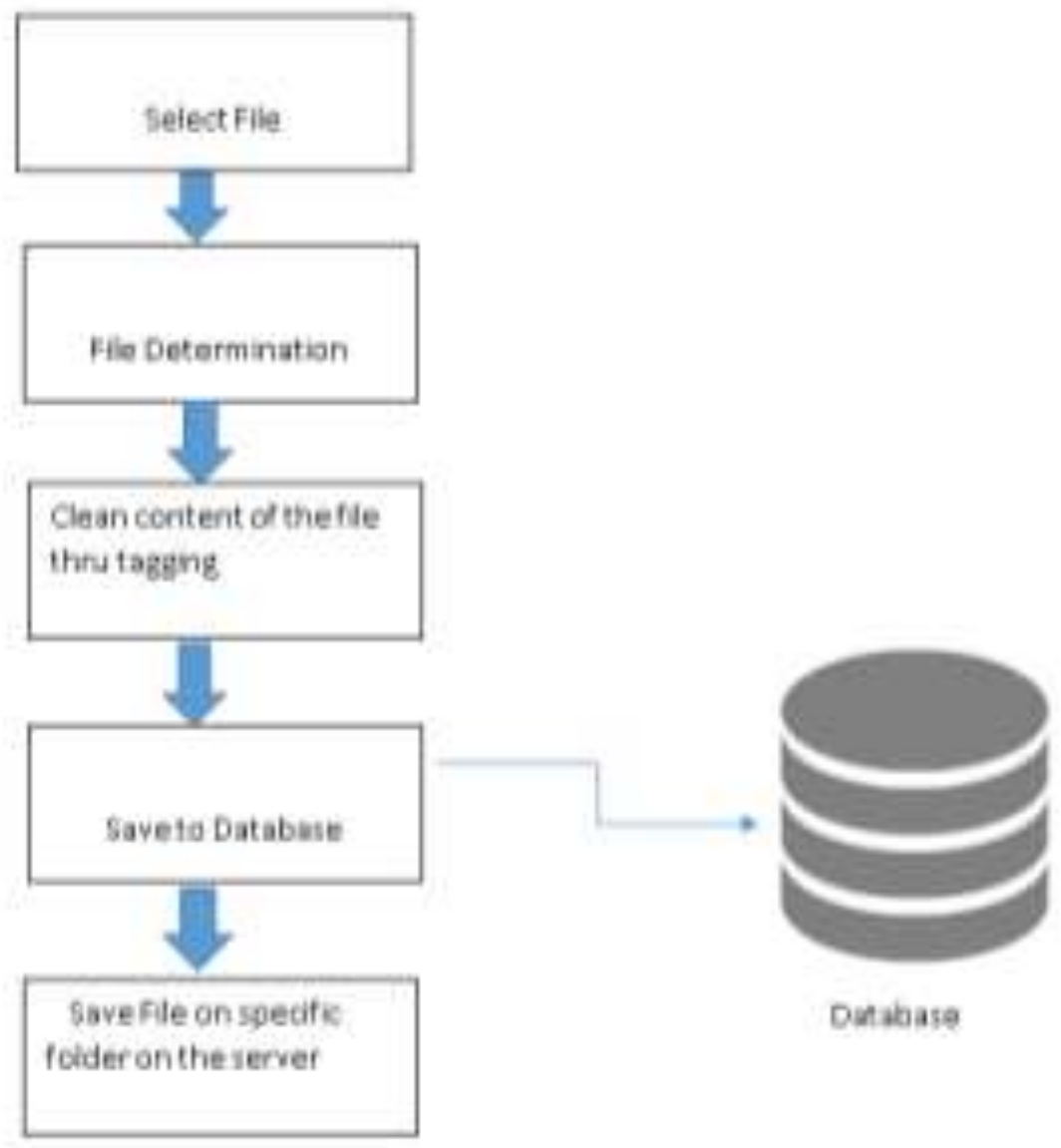

*Figure* 2. Process of indexing

The relevant documents which are stored in the database are clustered based on the given queries. For example, if the user searches the topic "data mining" the documents which are related to the data mining and also additional documents for the given topic are retrieved. All retrieved documents are clustered by considering similarities.

## RESULTS AND DISCUSSIONS

A total of 200 questionnaires were distributed with a return rate of 100%. The questionnaire was tested 0.735 using Cronbach's Alpha Coefficient and considered a reliable instrument. Evaluation is based on a Likert scale of Excellent, Good, Fair, and Poor as described for the selected quality characteristics. The questions asked for the quality characteristics as per in Table I. Data were processed and analyzed using SPSS (version 18) where the instrument was evaluated to test the reliability and validity of the measures used.

Table 1. Quality Characteristics Questions

| No | Questions |
|---|---|
| 1 | The suitability of the system. |
| 2 | The ability of the system to produce the expected result. |
| 3 | The ability of a system to handle the error. |
| 4 | The ability of a system to resume working in the event of an error. |
| 5 | The ability of users to understand the system. |
| 6 | The ability of users to learn the system easily. |
| 7 | The use of a suitable interface in the system. |
| 8 | The operation speed of the system. |

Four quality characteristics were evaluated in this study; Usability, Performance Efficiency, Reliability, and Functionality Suitability. Figure 3 illustrates results of the Usability evaluation where 15% of the respondents perceive the system as being "Excellent", 63% as being "Good", and 22% as being "Fair". None of the respondents found the applications as "Poor". These results are as expected although the high acceptance rate (more than 70%) came as a surprise. This indicates that the system is perceived as 'usable' by the teachers in aiding the learning process.

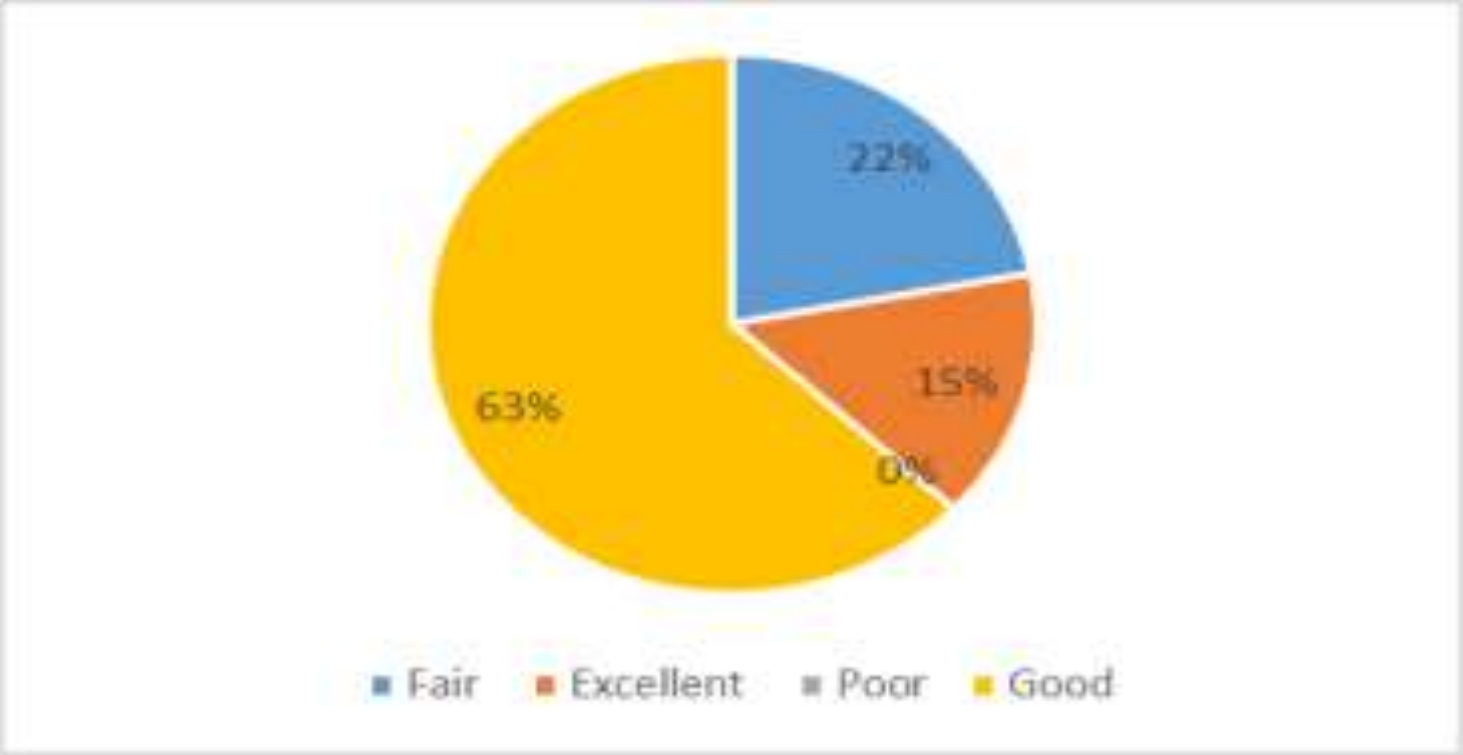

*Figure 3*. Evaluation of Usability

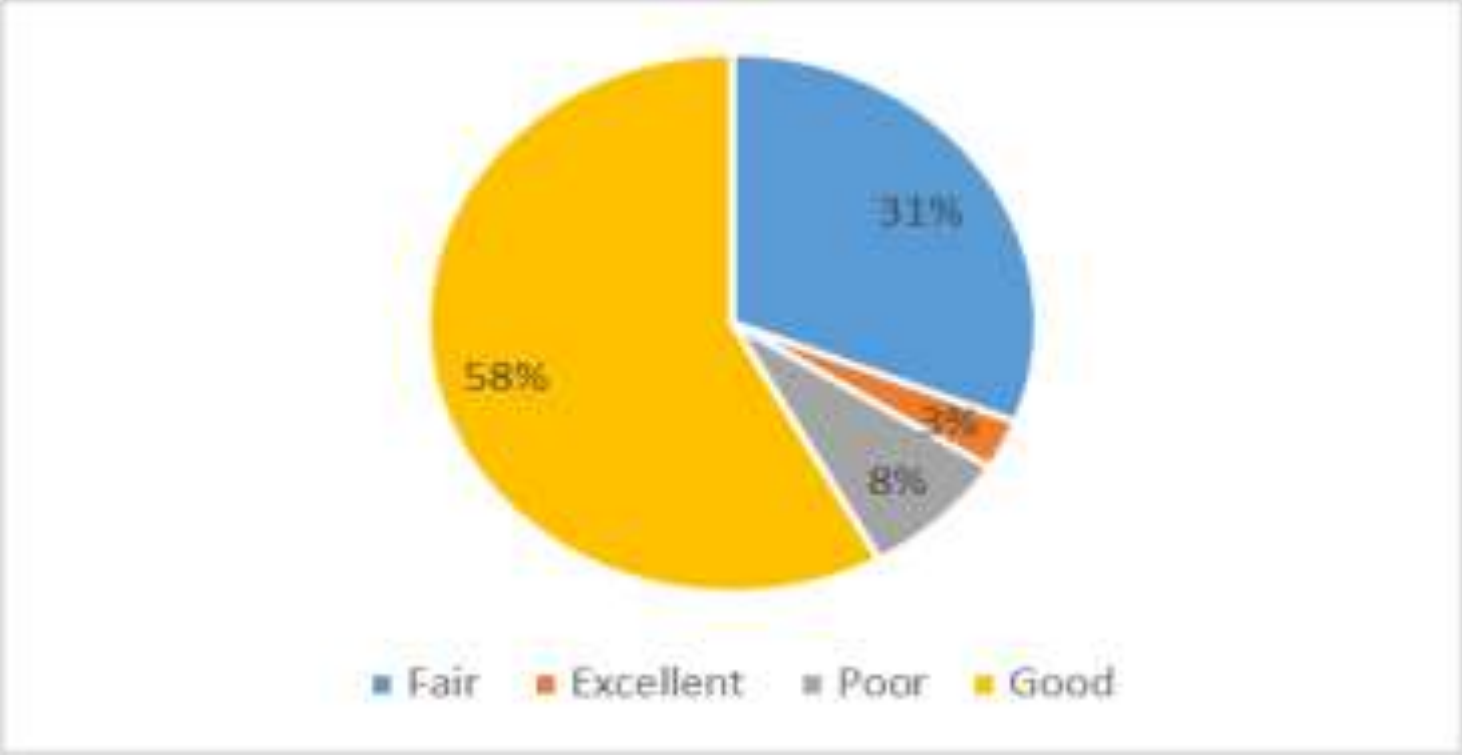

*Figure 4*. Evaluation of Reliability

Figure 4 illustrates results of the Reliability evaluation where 3% perceive the system as being ‘Excellent’, 30% as being 'Good", 31% as being 'Fair', and 8% as being 'Poor'. The 8% 'Poor' rating can be attributed to the need for more functions in the system. Nevertheless, based on the majority of respondents (78%), it can be said that the applications are ‘reliable’ for document clustering.

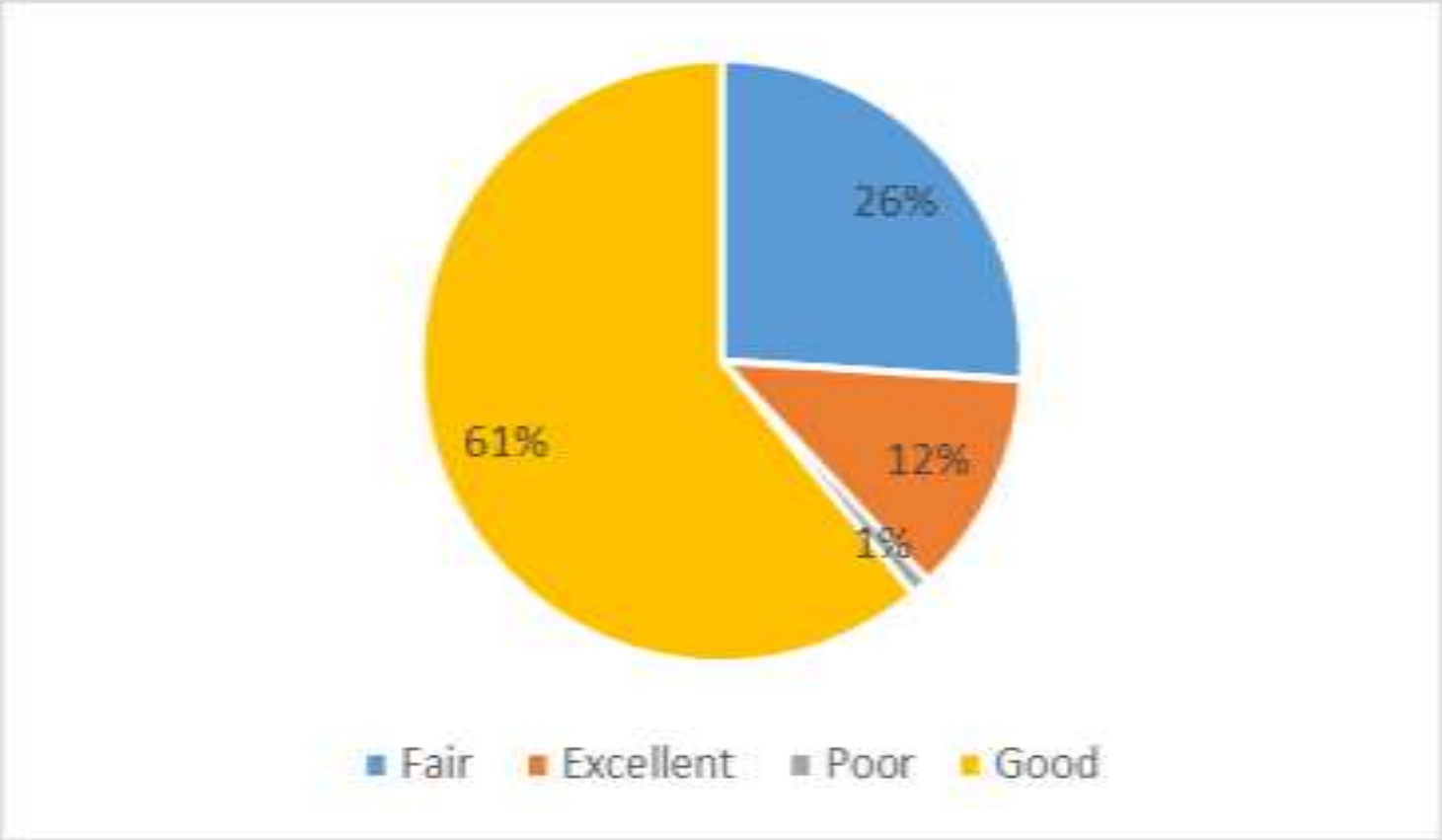

*Figure* 5. Evaluation of Functional Suitability

Results of the Functional Suitability evaluation are depicted in Figure 5 where 12% of the system as being 'Excellent', 61% as being ‘Good”, 26% as being ‘Fair', and 1% as being 'Poor'. We were expecting a much higher rate of acceptance although the overall results indicate that the system is perceived as 'functional' by the Mindoro State University administrator, teacher, and personnel.

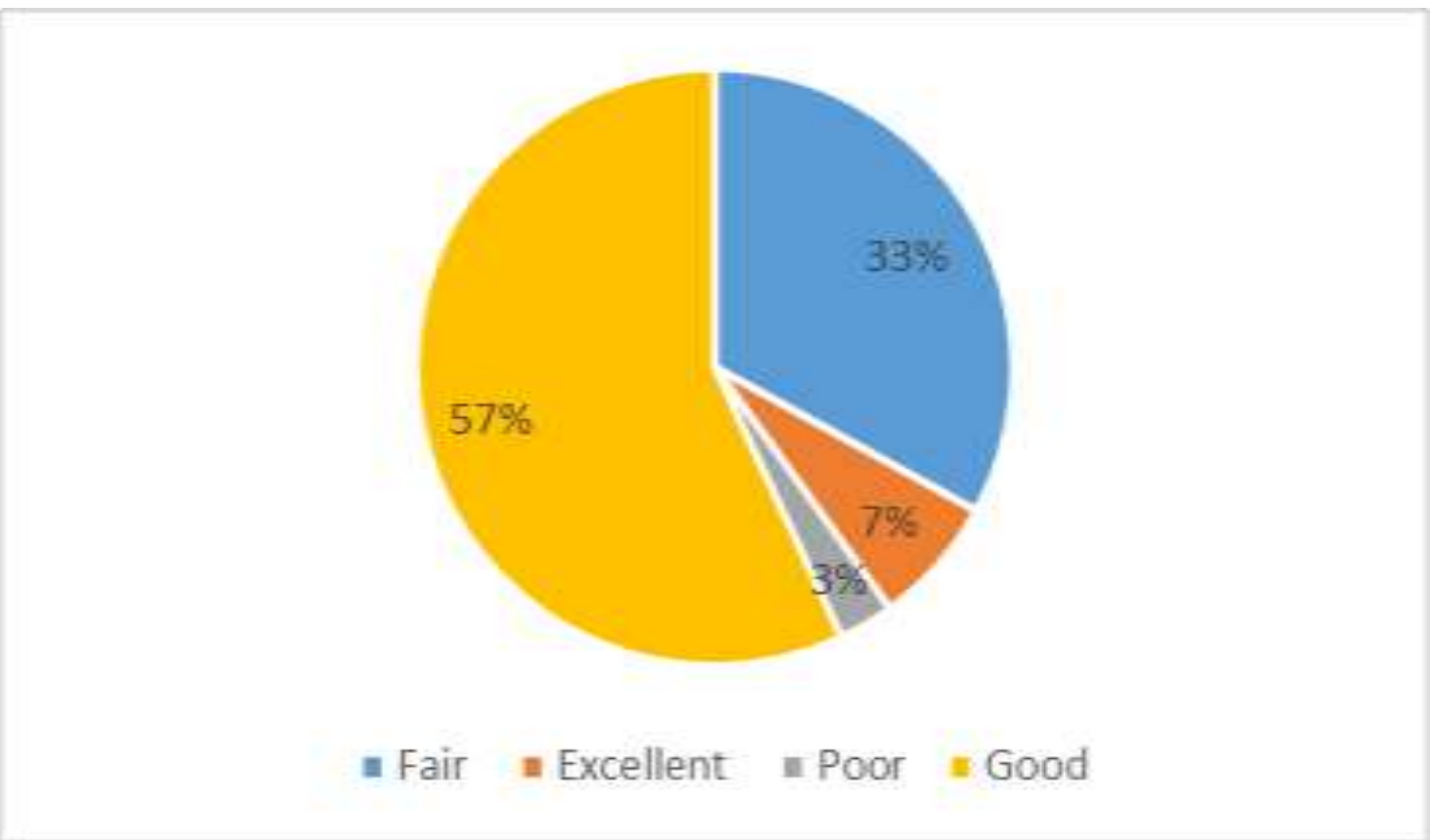

*Figure* 6. Evaluation of Performance Efficiency

Finally, the results of the Performance Efficiency evaluation are depicted in Figure 6where 7% perceive the system as being 'Excellent', 57% as being 'Good", 33% as being 'Fair', and 3% as being 'Poor'. We were expecting a lower rate of acceptance due to the hardware specifications of the system. However, the overall results indicate that developed systems are perceived as 'efficient' with 3% of the respondents feeling that the performance and hardware specifications need upgrading.

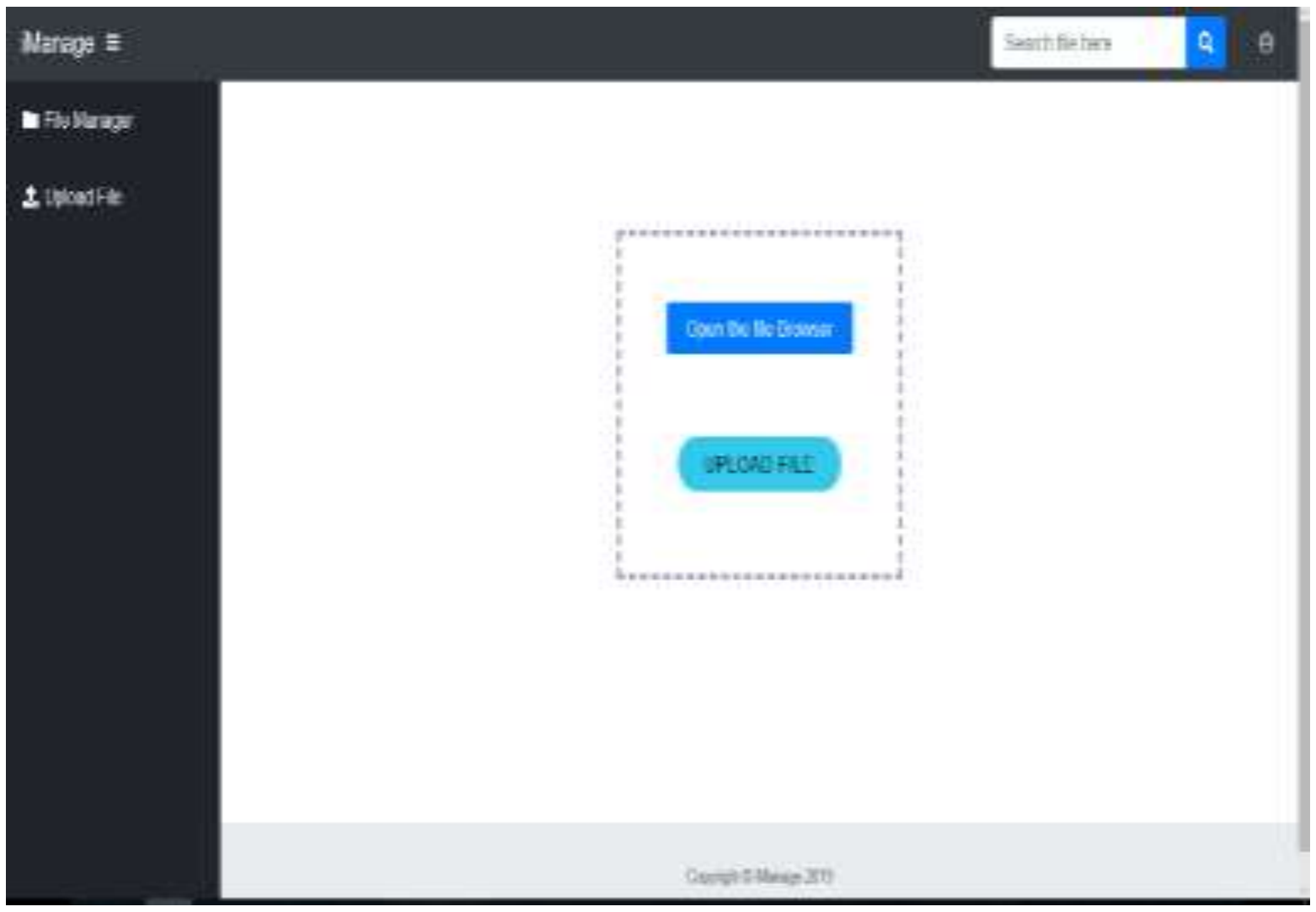

*Figure 7*. Screenshot of User Interface for Uploading of documents

As shown in Figure 7 the screenshot of the interface for easy uploading of documents. Initially, the user will select and upload a file from the collection of documents.

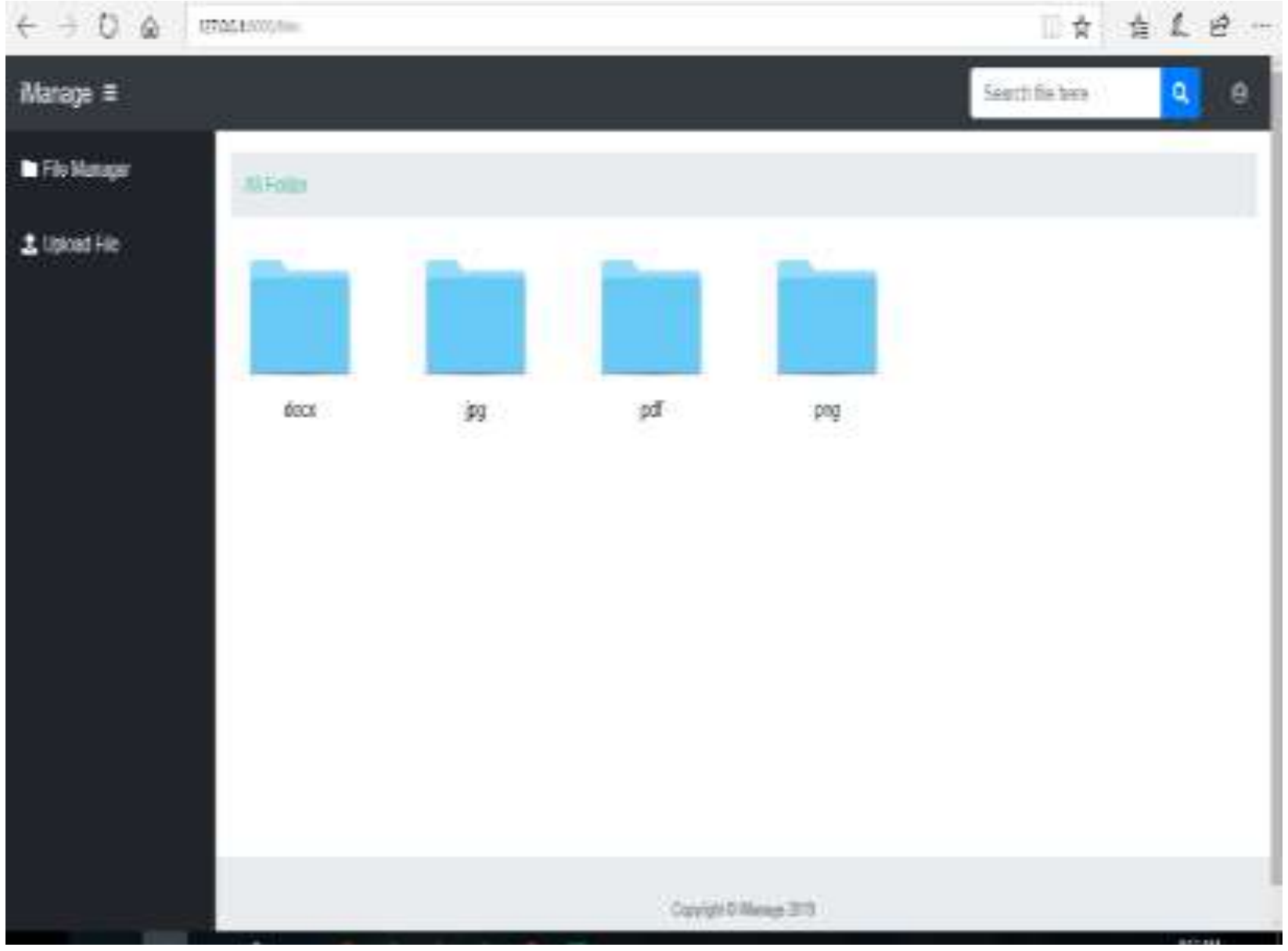

*Figure 8*. Screenshot of files clustering

Shown in Figure 8 is the clustering of files. The system determines and analyzes the uploaded file whether it is a file (word) document, pdf format, and photos/images.

The documents clustering algorithms attempt to group the document using similarity measures.

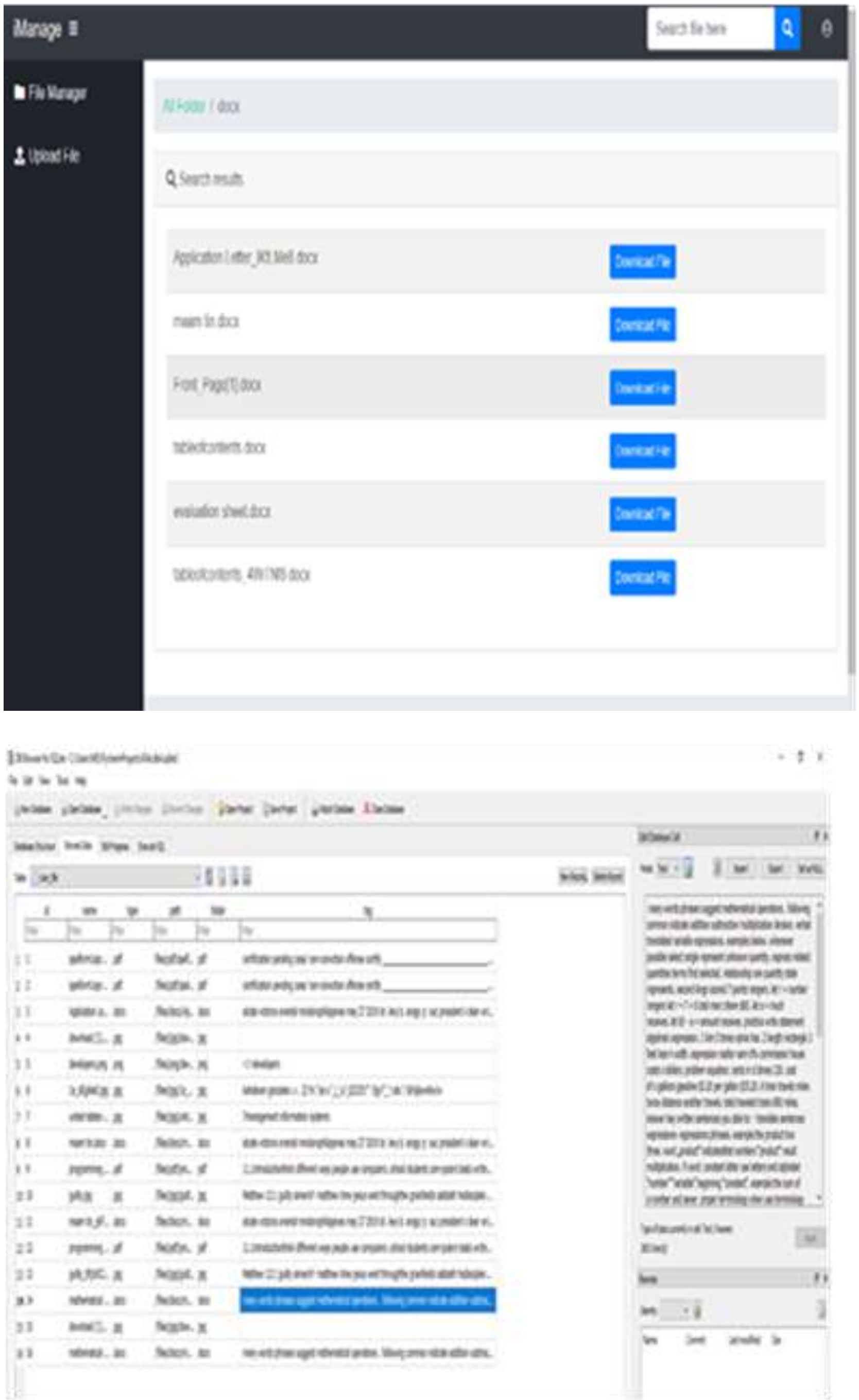

*Figure 9*. Summary of uploaded files and stop words

Shown in Figure 9 is the summary of uploaded files and stop words. Sometimes a very common word, which would appear to be of little significance in helping to select documents matching users' needs, is completely excluded from the vocabulary. These words are called "stop words" and the technique is called "stop word removal". The general strategy for determining a “stop list” is to sort the terms by collection frequency and then to make the most frequently used terms, as a stop list, the members of which are discarded during indexing. The stop words are used to eliminate the unwanted words such as before, is, a, an, the, become, then, they, there, that, them, etc.

## CONCLUSIONS AND RECOMMENDATIONS

Based on the results of the evaluation of the system, the Web-based file clustering could help in reaching the required documents that the user is searching for. Searching files is a task that consumes too much time and effort especially for ambiguity queries that have many meanings. Data clustering is used to improve the information retrieval from the collection of documents. A novel approach has been introduced for search results clustering that is based on the semantics of the retrieved documents rather than the syntax of the terms in those documents. The need for an information retrieval mechanism can only be supported if the document collection is organized into a meaningful structure, which allows part or all the document collection to be browsed at each stage of a search. After testing the system, it is recommended that upon uploading of file it will show the use of the file and where it is originated (department). It is also recommended to create an index to cluster not only the file type but also the content and use of a file. Explore the clustering to a wider scope.

## ACKNOWLEDGEMENT

The authors would like to acknowledge and gratitude the Administration of Mindoro State University, College of Computer Studies, and College of Engineering, Laguna University.